\documentclass[
aps,
prapplied,
reprint,
superscriptaddress,
longbibliography,
floatfix
]{revtex4-2}

\usepackage{graphicx}
\usepackage{amsmath}
\usepackage{amssymb}
\usepackage{braket}
\usepackage{dcolumn}
\usepackage{bm}

\begin{document}

\preprint{APS/123-QED}

\title{Quantum gyroscope based on three-dimensional rotation induced Berry phase}
\author{Huaijin Zhang$^{1,2}$ and Zhang-Qi Yin}
\email{zqyin@bit.edu.cn}
\affiliation{Key Laboratory of Advanced Optoelectronic Quantum Architecture and Measurements (MOE), and Center for Quantum Technology Research, School of Physics, Beijing Institute of Technology, Beijing 100081, China}
\affiliation{Beijing Key Laboratory of Quantum Matter State Control and Ultra-Precision Measurement Technology, School of Physics, Beijing Institute of Technology, Beijing 100081, China}

\date{\today}

\begin{abstract}
Solid-spin defects in diamond provide long coherence times and room-temperature optical initialization and readout, making them an attractive platform for compact solid-state quantum gyroscopes. A central challenge for NV-based gyroscopes is that the rotation-induced signal is weak, while near-resonant operation, although enhancing the response, can induce nonadiabatic transitions that degrade the accumulated geometric phase and readout fidelity. Here we investigate a levitated diamond under three-dimensional rotation, in which intrinsic ${}^{14}\mathrm{N}$ nuclear spins associated with NV centers act as sensing qubits. We show that the rotation is encoded in a geometric (Berry) phase and identify a near-resonant regime with strongly enhanced phase response. To suppress the resulting nonadiabatic leakage, we introduce a counter-diabatic protocol derived from the Kato gauge potential. This enables robust geometric-phase accumulation and improves the sensitivity by four orders of magnitude relative to the conventional detuned protocol. We further evaluate the achievable sensitivity and the dominant experimental limitations, including decoherence and protocol overhead, thereby establishing a realistic route toward high-performance NV-based solid-state quantum gyroscopes.
\end{abstract}

\maketitle

\section{Introduction}

High-precision rotation sensing is essential for inertial navigation and for experimental tests of fundamental physics. State-of-the-art gyroscopes are currently dominated by large-scale optical and atomic platforms, most notably ring-laser gyroscopes and atom interferometers, which achieve exceptional sensitivity through the optical or matter-wave Sagnac effect \cite{stedman1997ring,schreiber2013invited,cronin2009optics,gustavson1997precision,barrett2014sagnac}. Their performance, however, typically relies on bulky infrastructure, including extended optical paths, vacuum systems, and complex laser assemblies, which limits their deployment in compact and portable settings. These constraints have motivated growing interest in solid-state quantum platforms for rotation sensing, where high sensitivity could in principle be combined with miniaturization and integrability. Among solid-state candidates, nitrogen-vacancy (NV) centers in diamond are particularly appealing because they combine long spin coherence times with well-established optical initialization and readout at room temperature \cite{taylor2008high,balasubramanian2009ultralong,rondin2014magnetometry}.

Recent advances in levitated optomechanics and spin-mechanical platforms have made NV-based solid-state gyroscopes a far more realistic prospect \cite{millen2020optomechanics,gonzalez2021levitodynamics}. Stable high-speed rotation has been demonstrated for both levitated micro- and nanoparticles in vacuum \cite{hoang2016torsional,ahn2018optically,jin2024quantum,sha2025spin}, while levitated diamonds hosting NV centers have been realized and coherently controlled in both optical and electrodynamic traps \cite{delord2017electron,delord2018ramsey}. In parallel, feedback cooling and related optomechanical techniques now enable increasingly precise control of mechanical motion \cite{gieseler2012subkelvin,chang2010cavity,delic2020cooling,delord2020spin,liu2026cooling}, and mature microwave and radio-frequency pulse protocols allow high-fidelity manipulation of NV spins \cite{de2010universal,bar2013solid,childress2006coherent,london2013detecting}. Motivated by these developments, several NV-based gyroscope protocols have been proposed \cite{ledbetter2012gyroscopes,ajoy2012stable,Zhang2023highly,zeng2024optically,soshenko2021nuclear,jarmola2021demonstration}. These schemes generally rely on Ramsey interferometry, spin precession, or related frequency-shift measurements, in which the rotation signal is inferred from spin-phase accumulation. A central challenge is that the coupling between spin and mechanical rotation is intrinsically weak, often much smaller than the corresponding couplings to magnetic fields or strain \cite{maclaurin2012measurable,wood2017magnetic,teissier2014strain}. As a result, the accumulated phase grows slowly and the achievable sensitivity remains limited. A variety of enhancement strategies have therefore been explored, including strain-mediated coupling \cite{teissier2014strain,macquarrie2013mechanical}, hyperfine-assisted schemes \cite{wang2024hyperfine}, nuclear-spin ensembles \cite{ajoy2012stable}, and transverse driving fields \cite{loretz2013radio,stark2017narrow}. Despite these efforts, a substantial performance gap remains between solid-state gyroscopes and state-of-the-art optical or atomic devices \cite{stedman1997ring,schreiber2013invited,cronin2009optics,gustavson1997precision,barrett2014sagnac,ledbetter2012gyroscopes,ajoy2012stable}. This persistent gap points to the need for a more efficient mechanism for converting mechanical rotation into a measurable spin phase \cite{ledbetter2012gyroscopes,ajoy2012stable,maclaurin2012measurable,wood2017magnetic,teissier2014strain,macquarrie2013mechanical,loretz2013radio,stark2017narrow}.

Here we propose an NV-based gyroscope that exploits the geometric (Berry) phase of intrinsic $^{14}\mathrm{N}$ nuclear spins in a levitated diamond undergoing three-dimensional rotation. Geometric-phase-based sensing in mechanically rotated NV-diamond systems has recently emerged as a promising direction \cite{zhang2025ultra}, and, in the noninertial frame of the spin system, the external rotation $\Omega$ acts as an effective magnetic field, analogous to the Barnett effect \cite{barnett1915magnetization,wood2017magnetic}. This provides a route to encode the rotation rate in a geometric phase rather than relying solely on conventional dynamical-phase accumulation. A key feature of our protocol is a near-resonant regime in which the geometric-phase response is strongly amplified, especially near the avoided crossing, thereby enhancing the rotational scale factor. The difficulty is that this is also the regime in which adiabaticity is most susceptible to breakdown: as the energy gap decreases, nonadiabatic transitions become significant and degrade the signal \cite{berry1984quantal,kato1950adiabatic,demirplak2003adiabatic,demirplak2005assisted,berry2009transitionless}. To overcome this limitation, we introduce transitionless counter-diabatic control formulated in terms of the Kato gauge potential \cite{kato1950adiabatic,demirplak2003adiabatic,demirplak2005assisted,berry2009transitionless,guery2019shortcuts}, which suppresses nonadiabatic leakage while preserving the near-resonant enhancement. This leads to a four-orders-of-magnitude improvement in sensitivity relative to the conventional detuned protocol. We further evaluate the achievable sensitivity under realistic experimental conditions, taking into account decoherence and protocol overhead within the standard framework of quantum sensing \cite{degen2017quantum}. Our analysis indicates that an ensemble of $10^6$ NV centers can reach a shot-noise-limited sensitivity of approximately $0.6~\mu\mathrm{rad/s}/\sqrt{\mathrm{Hz}}$ at the optimal operating point.

The remainder of this paper is organized as follows. Section II establishes the Hamiltonian of the $^{14}\mathrm{N}$ nuclear spin under three-dimensional mechanical rotation and analyzes the corresponding spin dynamics. Section III derives the counter-diabatic driving field within the Kato gauge-potential formalism and evaluates the resulting geometric phase. Section IV quantifies the rotation sensitivity in the near-resonant regime. Section V examines the dominant noise sources and assesses the practical feasibility of the proposed scheme. Finally, Sec. VI summarizes the main results and discusses the experimental requirements for implementing the protocol.

\section{Model}

\begin{figure}[htbp]
    \centering
    \includegraphics[width=8.5cm]{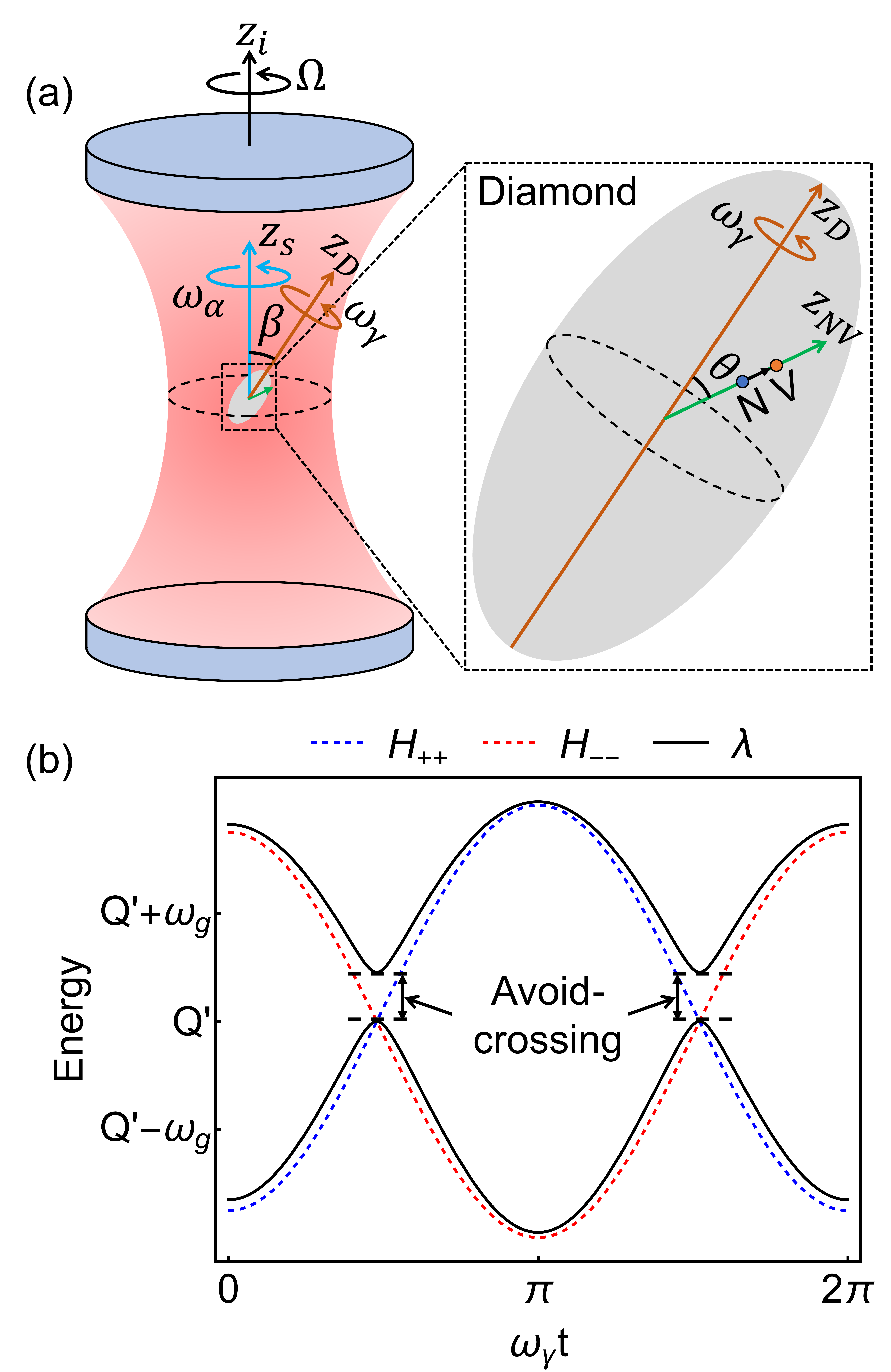}
    \caption{(a) Schematic illustration of the rotation sensing setup using a levitated micro-diamond. The diamond spins with frequency $\omega_\gamma$ about its principal axis $z_D$ and precesses with frequency $\omega_\alpha$ about the trap symmetry axis $z_S$ at a tilt angle $\beta$. The entire system undergoes an external rotation with frequency $\Omega$ about the inertial axis $z_i$, which is parallel to $z_S$. The axis of the nitrogen-vacancy (NV) center, $z_{\text{NV}}$, forms an angle $\theta$ with $z_D$. (b) Time evolution of the NV nuclear spin energy levels over one rotation period $T=2\pi/\omega_\gamma$. The dashed blue and red curves represent the diagonal Hamiltonian terms $H_{++}$ and $H_{--}$, respectively. The solid black curves $\lambda$ denote the instantaneous eigenvalues. The parameters are chosen to satisfy the resonance condition, leading to avoid-crossings at the intersection points of the diagonal terms.}
    \label{fig:1}
\end{figure}

We consider a levitated microdiamond embedded nitrogen-vacancy (NV) centers as the experimental platform for rotation sensing, as illustrated in Fig.~\ref{fig:1}(a). 
The diamond is optically trapped in a vacuum and undergoes controlled rotational motion, while the embedded NV center $^{14}\text{N}$ nuclear spin can be initialized and coherently controlled optically.
To describe the rotation sensing mechanism, we define four nested coordinate frames: the inertial frame ($z_i$), the system (or laboratory/trap) frame ($z_s$), the diamond principal-axis frame ($z_D$), and the NV-axis frame ($z_{NV}$). 
As illustrated in Fig.~\ref{fig:1}(a), the entire system undergoes a global rotation with frequency $\Omega$ along the $z_i$ axis.
This external rotation is represented by the operator $R_{z_i}(\Omega t) = \exp(-i \Omega t I_{z_i})$, where $I_z$ is the $^{14}\text{N}$ nuclear spin operator along the inertial axis. 
Within the system frame, the microdiamond undergoes 3D rotation parameterized by the Euler angles $\{\alpha, \beta, \gamma\}$.
We define the time-dependent rotation operator as $R(t) = R_{z_s}(\omega_\alpha t) R_{y_s}(\beta) R_{z_s}(\omega_\gamma t)$, under the assumption of negligible nutation ($\dot{\beta}=0$). 
The NV axis $z_{NV}$ is fixed relative to the diamond principal axis $z_D$ at an angle $\theta$, described by the transformation $R_{y_s}(\theta) = \exp(-i \theta I_{y_s})$.

We start from the bare spin Hamiltonian of the NV center \cite{doherty2013nitrogen}: $H_0=DS_z^2+QI_z^2+A_{\parallel}S_{z}I_{z}+(A_{\perp}/2)\left(S_{+}I_{-}+S_{-}I_{+}\right)$. 
where $D$ is the zero-field splitting of the electronic spin, $Q$ is the nuclear quadrupole coupling constant, and $A_{\parallel}$ and $A_{\perp}$ are the hyperfine coupling strengths.
When the electronic spin is initialized in the $m_s=0$, the transverse hyperfine interaction can be treated perturbatively, yielding the effective nuclear-spin Hamiltonian \cite{zhang2025ultra}: $H_N = Q' I_z^2$, with the revised quadrupole coupling $Q'=Q+A_{\perp}^2/ D$. 
To analyze the spin dynamical response to an external rotation $\Omega$, we perform a series of frame transformations. 
First, the Hamiltonian in the system frame $H_s$ is given by $H_s = \mathcal{R}(t) H_N \mathcal{R}^\dagger(t)$, where the total rotation $\mathcal{R}(t) = R(t)R_{y_s}(\theta)$ accounts for both the 3D mechanical motion of the diamond and the internal alignment of the NV center. 
In the inertial frame, the Hamiltonian is $H_i = R_{z_i}(\Omega t) H_s R_{z_i}^\dagger(\Omega t)$. 
To extract the inertial contribution of $\Omega$, we transform $H_i$ back to the system frame, yielding:
\begin{equation}
    H_s' = R_{z_i}^\dagger(\Omega t) H_i R_{z_i}(\Omega t) + i \dot{R}_{z_i}^\dagger(\Omega t) R_{z_i}(\Omega t) = H_s - \Omega I_{z_s},
\end{equation}
where the term $-\Omega I_{z_s}$ emerges as an effective inertial magnetic field directed along the system rotation axis. 
This is a direct consequence of the non-inertial nature of the rotating frame \cite{maclaurin2012measurable,wood2017magnetic}.
Finally, to facilitate the calculation of the geometric phase, we move into the co-rotating NV-axis frame via the transformation $W(t) = [\mathcal{R}(t)]^\dagger$ \cite{zhang2025ultra}. 
The total effective Hamiltonian in the NV body-fixed frame is:
\begin{equation}
    H = W H_s' W^\dagger + i (\partial_t W) W^\dagger.
    \label{eq:total_H}
\end{equation}
By substituting the explicit forms of $R(t)$ and $R_{y_s}(\theta)$, the Hamiltonian takes the form $H(t) = Q' I_z^2 - \gamma_n\vec{B}_{\text{eff}}(t) \cdot \vec{I}$. 
Here, the effective magnetic field $\vec{B}_{\text{eff}}$ contains contributions from the intrinsic rotations ($\omega_\alpha, \omega_\gamma$) and the external signal $\Omega$. 
Crucially, the external signal $\Omega$ is coupled to the spin degrees of freedom through the transformation $W(t)$, establishing the fundamental sensing mechanism for the proposed gyroscope.

To reveal the physical origin of the enhanced rotational sensitivity, we investigate the instantaneous eigenstates and eigenvalues of $H(t)$. 
In the standard spin-1 basis $\{|+1\rangle, |0\rangle, |-1\rangle\}$, the diagonal matrix elements $H_{++}$ and $H_{--}$ correspond to the unperturbed energy levels of the spin states $|+1\rangle$ and $|-1\rangle$, respectively. 
The physical basis for the sensing scheme is dynamically captured in Fig.~\ref{fig:1}(b). 
The occurrence of a resonance is determined by whether these original energy levels $H_{++}$ and $H_{--}$ intersect during a rotation period. 
If they cross, the system enters a resonance regime. 
At these intersection points, the transverse components of the effective field $\vec{B}_{\text{eff}}(t)$ induce a coherent coupling between the two states, causing the instantaneous eigenvalues of the full Hamiltonian to exhibit an avoid-crossing \cite{zhang2025ultra}. 

\begin{figure}[htbp]
    \centering
    \includegraphics[width=8.5cm]{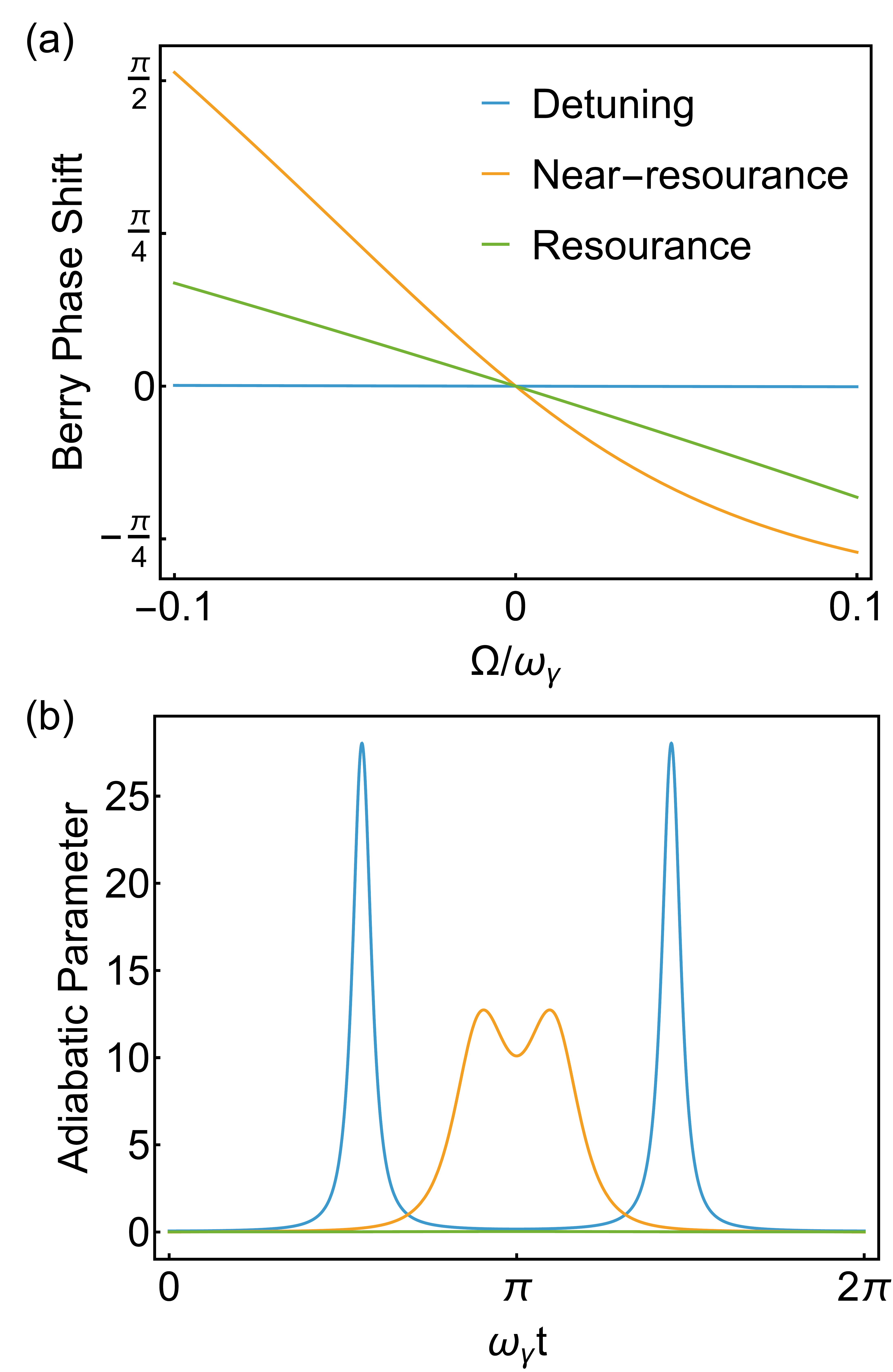}
    \caption{(a) The Berry phase shift (relative to the phase at $\Omega=0$) as a function of the normalized system rotation frequency $\Omega/\omega_\gamma$. Three distinct regimes are presented: large detuning (blue line), near-resonance (orange line), and full resonance (green line). The phase shift in the large detuning regime remains negligible, while the near-resonance case exhibits the most significant sensitivity to $\Omega$. (b) Time evolution of the adiabatic parameter over one period. The adiabatic condition requires this parameter to be much less than $1$. The results indicate that the adiabatic condition breaks down for the full resonance and near-resonance regimes (where the parameter exceeds $1$), leading to a discrepancy between the theoretical Berry phase and the actual accumulated geometric phase.}
    \label{fig:2}
\end{figure}

In the immediate vicinity of an avoid-crossing, the instantaneous eigenstates undergo a rapid and dramatic transformation. 
This abrupt change imparts an exceptional susceptibility to the eigenstates and the resultant geometric phase with respect to minute perturbations in external parameters (the global rotation frequency $\Omega$). 
The onset of this high-sensitivity resonance requires the original energy difference between the $\ket{+1}$ and $\ket{-1}$ states to vanish within the modulation cycle.
Formally, this necessitates that the extremum values of the original energy gap exhibit opposite signs over a single period $T=2\pi/\omega_\gamma$, leading to the following rigorous resonance condition \cite{zhang2025ultra}:
\begin{equation}
    \label{eq3}
    \left[\frac{\omega'_\alpha}{\omega_\gamma}\cos{(\theta+\beta)}+\cos{	\theta}\right]\left[\frac{\omega'_\alpha}{\omega_\gamma}\cos{(\theta-\beta)}+ \cos{\theta}\right]\leq0,
\end{equation}
where $\omega'_\alpha = \omega_\alpha + \Omega$ represents the effective precession frequency as modified by the external inertial rotation.

For a cyclic evolution in parameter space, the Berry-phase shift $\Delta\Phi_g$ relative to the $\Omega=0$ reference provides a direct probe of the external rotation $\Omega$ \cite{berry1984quantal,zhang2025ultra}. 
Figure~\ref{fig:2}(a) shows the calculated dependence of $\Delta\Phi_g$ across different regimes. 
In the large-detuning regime (blue curve), the geometric phase is only weakly affected by $\Omega$, resulting in a negligible phase response. 
By contrast, once the resonance condition is satisfied (green curve), the dependence of $\Delta\Phi_g$ on $\Omega$ is markedly enhanced. 
More importantly, as the system approaches the boundary of the resonance condition, i.e.,
\begin{equation}
    \omega_\alpha' \cos(\theta-\beta) \approx -\omega_\gamma  \cos	\theta,
\end{equation}
the phase slope $d\Phi_g/d\Omega$ increases sharply, as illustrated by the orange curve. 
This steep-response region is particularly favorable for rotation sensing, since even a minute variation in $\Omega$ can generate a pronounced geometric-phase signal.

Despite the strong phase response achievable near resonance, this regime is typically inaccessible in practice because the adiabatic approximation breaks down in the vicinity of the avoid-crossings. 
The corresponding adiabatic parameter is defined as
\begin{equation}
    \epsilon_{mn} = \left| \frac{\bra{\lambda_m} \dot{H}_{In} \ket{\lambda_n}}{(\lambda_n - \lambda_m)^2} \right|,
\end{equation}
with $\ket{\lambda_m}$ and $\ket{\lambda_n}$ the instantaneous eigenstates of the Hamiltonian.
As shown in Fig.~\ref{fig:2}(b), $\epsilon_{mn}$ remains far below unity in the detuned regime but increases sharply near resonance, where the avoid-crossing gap becomes minimal. 
The resulting nonadiabatic transitions degrade the fidelity of geometric-phase accumulation and therefore suppress the sensing advantage of resonant operation \cite{berry2009transitionless}. 
Consequently, previous schemes were limited either to off-resonant conditions with reduced sensitivity or to very slow driving protocols designed to maintain adiabatic evolution \cite{zhang2025ultra}. 
Here, we overcome this limitation by incorporating counter-diabatic driving, which suppresses nonadiabatic transitions and enables high-sensitivity operation directly in the resonant regime.

\section{Suppressing Nonadiabatic Transitions via Kato gauge potential}

\begin{figure}[htbp]
    \centering
    \includegraphics[width=8.5cm]{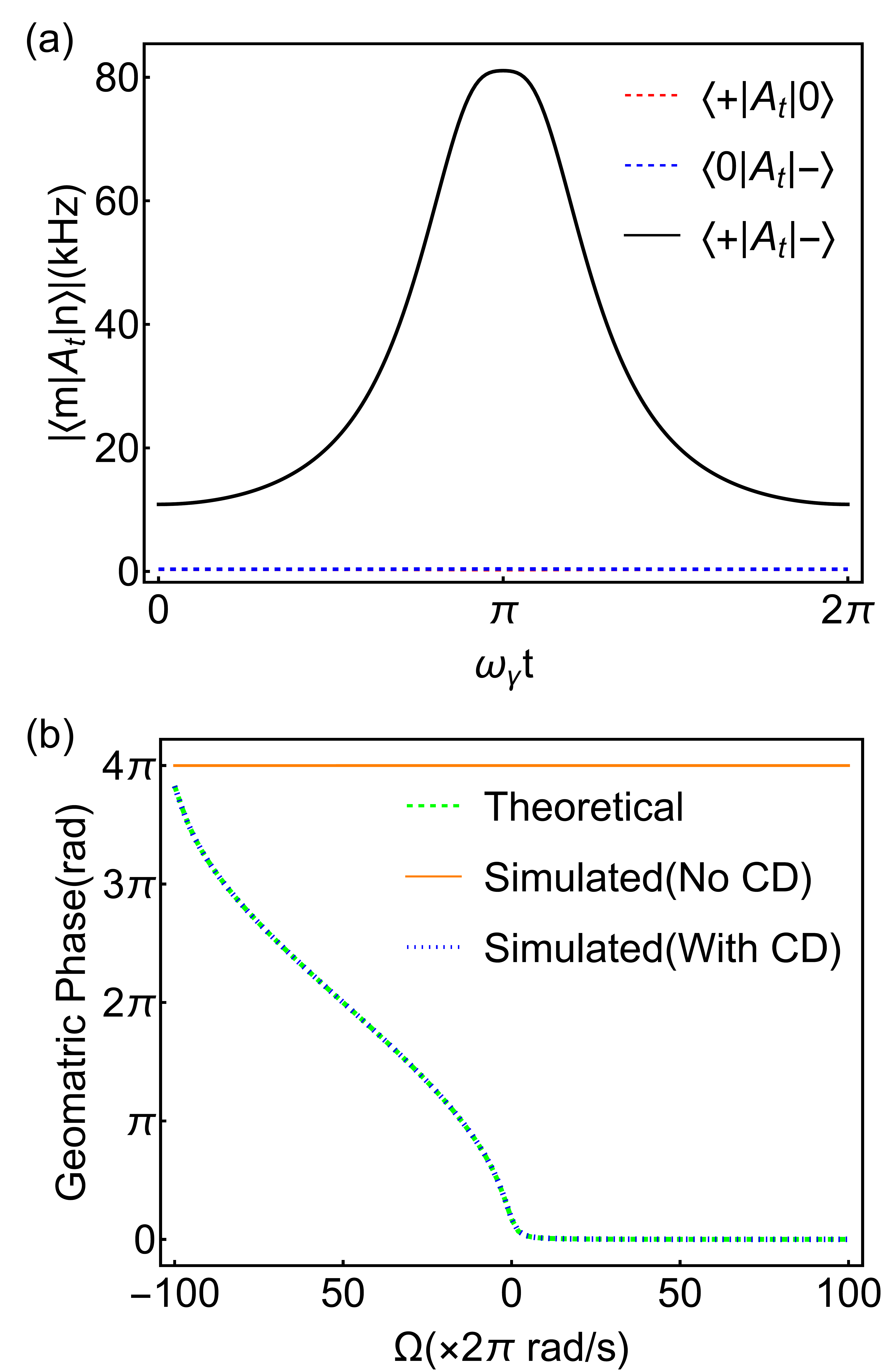}
    \caption{(a) Time dependence of the matrix elements of the required CD Hamiltonian (derived from the Kato gauge potential). The CD field induces a strong coupling between the $|+1\rangle$ and $|-1\rangle$ states (black solid line) with a peak amplitude on the order of $80$ kHz, while the couplings involving the $|0\rangle$ state (red and blue dashed lines) remain negligible. (b) The geometric phase as a function of the rotation rate $\Omega$ in the near-resonance regime. The green dashed line shows the theoretical prediction based on ideal adiabatic evolution. The orange solid line represents the numerical simulation without the CD term ("No CD"), which fails to follow the theoretical curve due to non-adiabatic transitions. The blue dotted line shows the simulation with the CD term ("With CD"), demonstrating that the inclusion of the Kato gauge potential effectively suppresses non-adiabatic errors and restores the expected geometric phase.}
    \label{fig:3}
\end{figure}

As established above, the near-resonant regime provides the largest geometric-phase response to the external rotation rate $\Omega$. 
Its practical use, however, is limited by strong nonadiabatic transitions that emerge in the vicinity of the avoid-crossing. 
These transitions prevent faithful adiabatic following of the instantaneous eigenstates, thereby reducing the accumulated geometric phase and compromising the measurement fidelity. 
To overcome this limitation without abandoning the sensitivity enhancement associated with near-resonant operation, we adopt transitionless quantum driving (TQD) and introduce the corresponding counter-diabatic (CD) Hamiltonian based on Kato gauge potential.

Within Kato gauge potential, the counter-diabatic term is uniquely constructed to enforce exact evolution along the instantaneous eigenstates $\ket{\lambda_n}$ of the reference Hamiltonian $H(t)$. 
It is written as \cite{kato1950adiabatic}:
\begin{equation}
    H_{\text{CD}}(t) = \frac{i}{2} \sum_n \left[ \partial_t(|\lambda_n\rangle\langle \lambda_n|), |\lambda_n\rangle\langle \lambda_n| \right].
    \label{eq:kato_potential}
\end{equation}
so that the corrected total Hamiltonian are generated by \begin{equation}
    H_{\text{tot}}(t) = H(t) + H_{\text{CD}}(t). 
\end{equation}

To clarify the control requirements implied by the counter-diabatic protocol, we examine the matrix elements of the analytically obtained Kato gauge potential in the bare spin basis $\{|+1\rangle, |0\rangle, |-1\rangle\}$. 
As shown in Fig.~\ref{fig:3}(a), the dominant contribution to $H_{\text{CD}}$ is a direct, time-dependent coupling between the $|+1\rangle$ and $|-1\rangle$ states (black solid curve). 
This coupling reaches a sharp maximum of approximately $80$ kHz at $\omega_\gamma t = \pi$, precisely where the avoid-crossing occurs and the probability of nonadiabatic Landau-Zener transitions is highest. 
In contrast, the matrix elements involving the $|0\rangle$ state, namely $|\langle +1|H_{\text{CD}}|0\rangle|$ and $|\langle 0|H_{\text{CD}}|-1\rangle|$ (red and blue dashed curves), emain negligibly small throughout the evolution. 
Therefore, the essential requirement for implementing the CD protocol is the realization of an effective transverse drive acting within the $\{|+1\rangle, |-1\rangle\}$ subspace.

The essential role of the Kato gauge potential is clearly demonstrated in Fig.~\ref{fig:3}(b), which shows the accumulated geometric phase $\Phi_g$ as a function of the external rotation rate $\Omega$ for operation in the near-resonant regime. 
The theoretical adiabatic prediction (green dashed curve) represents the ideal limit of perfect adiabatic following and exhibits a steep dispersive dependence on $\Omega$ in the vicinity of the zero point. 
By contrast, direct numerical simulations of the bare dynamics in the absence of the CD term (orange solid curve) show that this phase response is almost entirely lost. 
Strong nonadiabatic transitions at the avoid-crossing prevent the system from accumulating the intended geometric phase, leading to an essentially flat dependence of $\Phi_g$ on $\Omega$.

By contrast, once the counterdiabatic term $H_{\text{CD}}(t)$ is incorporated into the evolution, the nonadiabatic excitations are strongly suppressed, as shown by the blue dotted curve. 
The resulting simulated geometric phase closely follows the theoretical adiabatic prediction over the entire relevant range of $\Omega$. 
This recovery of the phase response confirms that the applied CD control realizes an effective shortcut to adiabaticity (STA). 
Consequently, the sensor can operate in the near-resonant sweet spot, where the geometric-phase response is maximized, while retaining high-fidelity phase accumulation. 
These results provide strong support for the theoretical feasibility of the proposed rotation-sensing scheme.

\section{measurement sensitivity}

\begin{figure}[htbp]
    \centering
    \includegraphics[width=8.5cm]{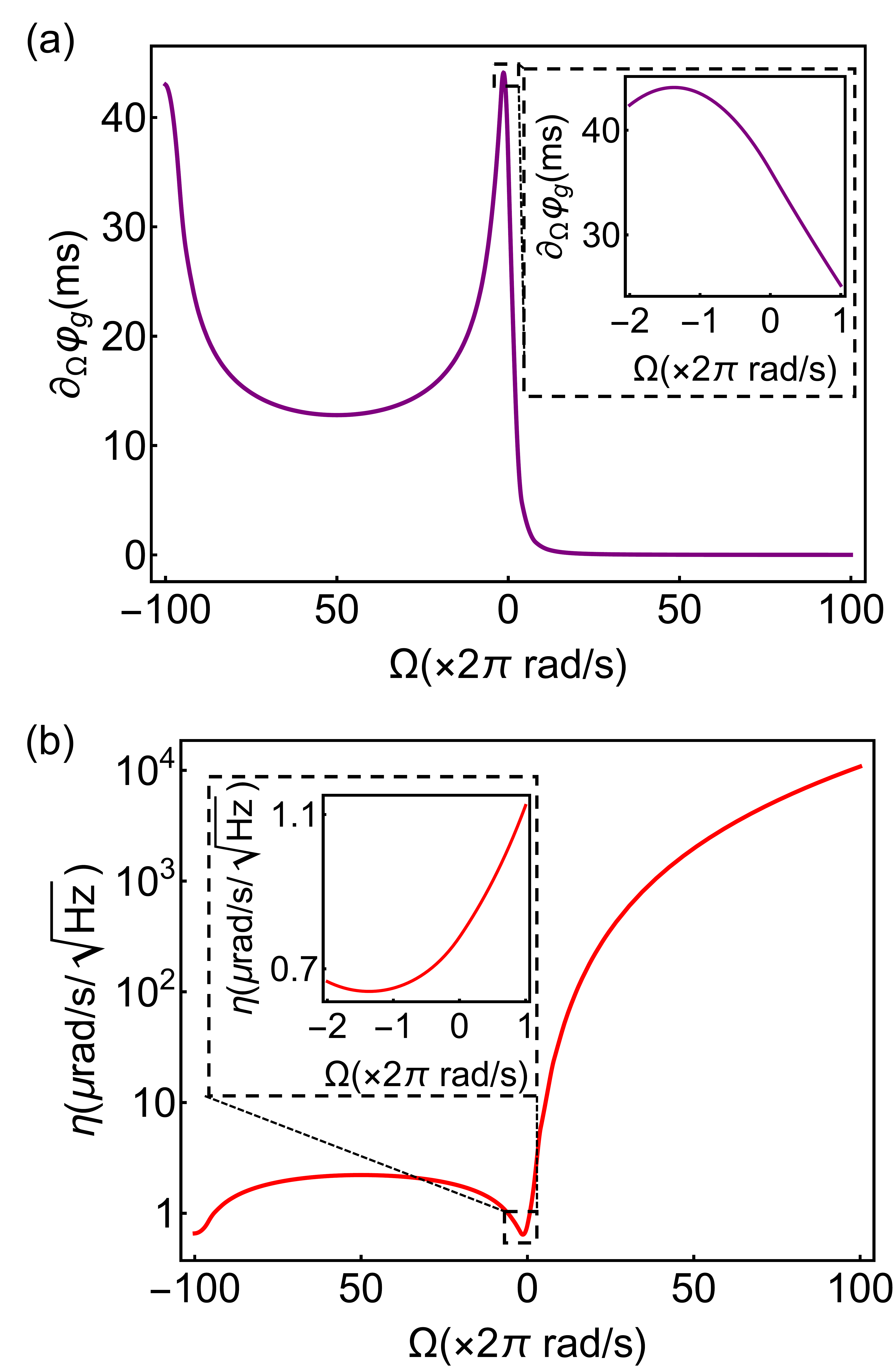}
    \caption{(a) The slope of the geometric phase with respect to the rotation frequency, defined as $\partial_\Omega \varphi_g(s)$, plotted as a function of $\Omega$. The steep slope near $\Omega \approx 0$ indicates a high sensitivity to rotation. The inset shows the zoomed-in view within the practical measurement bandwidth (approximately $\pm 1$ Hz), where the response remains smooth. (b) The theoretical sensitivity $\eta$ of the rotation measurement versus $\Omega$. The calculation assumes an ensemble of $N=10^6$ NV centers. The system achieves a shot-noise-limited sensitivity of approximately $0.6 \, \mu\text{rad/s}/\sqrt{\text{Hz}}$ at the optimal working point. The inset highlights the sensitivity in units of $\mu	\text{rad/s}/\sqrt{	\text{Hz}}$ in the weak transverse field regime.}
    \label{fig:4}
\end{figure}

Having established robust geometric-phase accumulation using Kato gauge potential, we now assess the theoretical performance of the proposed rotation-sensing scheme. 
For a practical gyroscope, the central figure of merit is the measurement sensitivity, which is determined by how strongly the measured observable (here the geometric phase $\varphi_g$) responds to the parameter of interest, namely the global rotation rate $\Omega$.

To characterize the intrinsic scale factor of the proposed sensor, we consider the rotational response coefficient $\partial_\Omega \varphi_g$, namely the derivative of the geometric phase with respect to the rotation frequency.
Figure~\ref{fig:4}(a) shows that this quantity is strongly amplified in the near-resonant regime. 
In particular, the sharp feature near $\Omega \approx 0$ originates from the rapid evolution of the instantaneous eigenstates around the avoid-crossing, leading to a greatly enhanced phase response to small rotational perturbations. 
The inset displays the corresponding behavior over a practically relevant bandwidth of approximately $\pm 1$ Hz. 
Notably, within this range the response remains continuous and smooth, which is essential for stable sensor operation and unambiguous phase-to-rotation conversion. 

It is also useful to contrast the present near-resonant geometric-phase protocol with a conventional nonresonant geometric-phase operation far from the avoid-crossing. 
In the latter case, the geometric phase depends only weakly on the rotation rate, and the corresponding response coefficient $|\partial_{\Omega}\varphi_g|$ is strongly suppressed. 
This behavior is directly visible in Fig.~\ref{fig:4} (a): away from the avoided-crossing region the phase slope approaches zero, implying negligible phase-to-rotation conversion and therefore poor sensitivity. 
By operating near the avoided crossing, the rapid evolution of the instantaneous eigenstates yields a strongly amplified geometric-phase slope, reaching $|\partial_{\Omega}\varphi_g|\sim 40$ in our parameters [Fig.~~\ref{fig:4} (a)], which directly enables the enhanced sensitivity reported below.

To quantify the achievable performance, we define the rotation sensitivity $\eta$, as the minimum detectable rotation rate referred to a $1$ s integration time. 
Assuming optimal interferometric readout, the uncertainty in the estimated rotation rate is set by the phase uncertainty divided by the response coefficient $|\partial_\Omega \varphi_g|$.
Assuming optimal interferometric readout, the shot-noise-limited sensitivity is given by \cite{budker2007optical}: 
\begin{equation}
    \eta =\delta \Omega\sqrt{T_m}=\frac{1}{|\partial_\Omega \varphi_g| \sqrt{N T_m}}
\end{equation}
where $N$ is the number of sensing spins and $T_m$ denotes the measurement time.

Figure~\ref{fig:4}(b) presents the theoretical rotation sensitivity $\eta$ as a function of the external rotation rate $\Omega$. 
For a realistic assessment of this solid-state sensing architecture, we consider an ensemble of $N=10^6$ active $^{14}\mathrm{N}$ nuclear spins, consistent with densities achievable in contemporary microdiamond platforms. 
Because the phase slope is strongly enhanced near the avoided crossing, the sensitivity is significantly improved in the weak transverse-field regime, namely when $\omega_\alpha\approx-\omega_\gamma$ and $\theta,\beta\ll1$. 
As shown in the inset of Fig.~\ref{fig:4}(b), the shot-noise-limited sensitivity reaches approximately $0.6~\mu\mathrm{rad/s}/\sqrt{\mathrm{Hz}}$ at the optimal operating point, representing an improvement of about four orders of magnitude over the detuned regime ($\Omega\gtrsim100~\mathrm{Hz}$).

To facilitate comparison with other NV-based gyroscope protocols, we further define the single-NV sensitivity as $\tilde{\eta}=\eta\sqrt{N}$. 
For the present protocol, this yields $\tilde{\eta}=6	\times 10^{-4}~\mathrm{rad/s}/\sqrt{\mathrm{Hz}}$. 
Using the same normalization, the schemes proposed in Refs.~\cite{ajoy2012stable,wang2024hyperfine} correspond to $\tilde{\eta}=3.45	\times 10^{1}~\mathrm{rad/s}/\sqrt{\mathrm{Hz}}$ and $\tilde{\eta}=5.52	\times 10^{-3}~\mathrm{rad/s}/\sqrt{\mathrm{Hz}}$, respectively. 
Therefore, our protocol achieves an approximately $10^{5}$-fold improvement in per-spin sensitivity over Ref.~\cite{ajoy2012stable} and about one order of magnitude improvement over Ref.~\cite{wang2024hyperfine}. 
The achieved sub-$\mu\mathrm{rad/s}/\sqrt{\mathrm{Hz}}$ sensitivity indicates that geometric-phase sensing near an avoided crossing, when stabilized by the Kato gauge potential, can enable NV-based microgyroscopes to approach performance levels competitive with state-of-the-art atomic interferometers, highlighting the potential of compact solid-state platforms for high-precision inertial sensing and navigation.

Although the above analysis is formulated for the $^{14}\mathrm{N}$ nuclear spin, the underlying sensing principle is not restricted to the nuclear-spin manifold and can, in principle, also be extended to the NV electronic spin. 
In that case, the relevant intrinsic energy scale is the electronic zero-field splitting $D\approx2.87~\mathrm{GHz}$, which is much larger than the nuclear quadrupolar splitting $Q\approx4.95~\mathrm{MHz}$. 
Since $D\gg Q$, one may generally expect a substantially stronger response for the electron-spin implementation under otherwise comparable conditions, suggesting that the rotation sensitivity could be further improved by roughly $2$--$3$ orders of magnitude. 
At the same time, the actual gain will depend on the specific operating point as well as the coherence properties, control fidelity, and readout efficiency of the electron-spin platform.

Moreover, although the present discussion focuses on rotation sensing, the same protocol can also be adapted for magnetometry, because rotation and magnetic field enter the spin Hamiltonian in an equivalent manner in the rotating frame. 
The corresponding magnetic-field sensitivity can be expressed as $\eta_B=\eta/\gamma$, where $\gamma$ is the gyromagnetic ratio. 
For the $^{14}\mathrm{N}$ nuclear-spin implementation, using $\gamma_n\approx1.93\times10^{7}~\mathrm{rad\,s^{-1}\,T^{-1}}$, we obtain
the corresponding megnetic field sensitivity $\eta_B^{(n)}\approx 31~\mathrm{fT}/\sqrt{\mathrm{Hz}}$. 
For an electron-spin realization, the magnetic-field sensitivity is expected to be further improved by several orders of magnitude owing to the much larger electronic gyromagnetic ratio, in addition to the larger intrinsic frequency scale discussed above.

\section{Noise analysis}

Although the shot-noise-limited sensitivity obtained above represents the ideal performance bound of the proposed rotation-sensing scheme, realistic solid-state operation is necessarily constrained by decoherence and experimental overhead. 
To evaluate the practical feasibility of the protocol, we now develop a noise model that includes the finite coherence time of the spin ensemble as well as the additional temporal costs introduced by initialization, control, and readout \cite{barry2020sensitivity,zhang2025ultra}.

In the nitrogen-vacancy (NV) platform, the sensor performance is primarily limited by two decoherence channels: longitudinal relaxation of the electron spin, characterized by the time constant $T_{1e}$, and transverse dephasing of the nuclear-spin ensemble, described by $T_{2n}^*$. 
Because the sensing protocol relies on the coupled dynamics of these two subsystems, the signal amplitude decays exponentially during the measurement time $T_m$. 
This combined effect can be captured by an effective coherence time,
\begin{equation}
    \tau = \frac{T_{1e}T_{2n}^*}{T_{1e}+T_{2n}^*}.
\end{equation}
Accordingly, the accumulated geometric-phase signal is suppressed by a factor of $\exp(-T_m/\tau)$, which in turn increases the measurement uncertainty.

\begin{figure}[htbp]
    \centering
    \includegraphics[width=8.5cm]{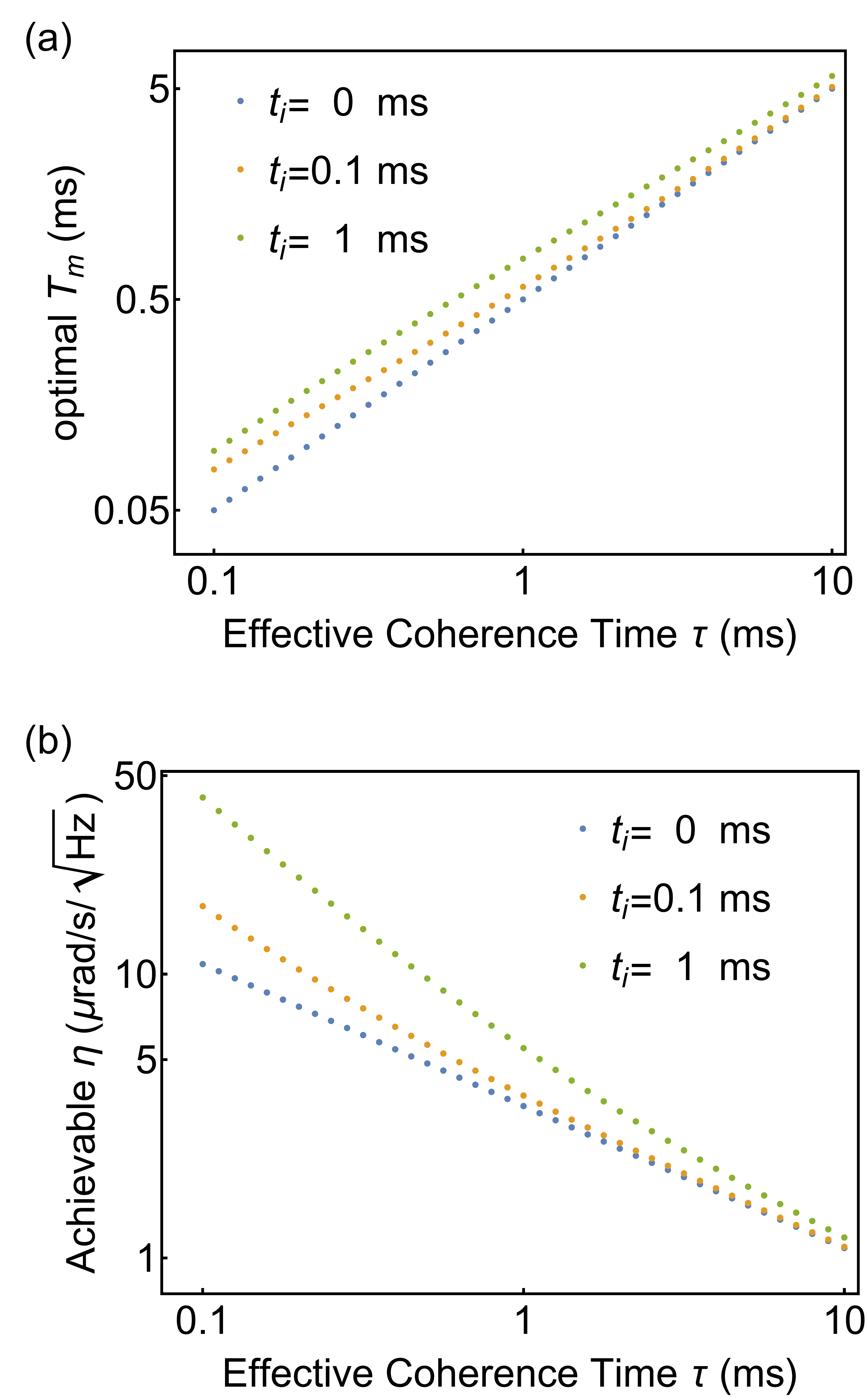}
    \caption{(a) The optimal measurement Time $T_m$ plotted as a function of the effective coherence time $\tau$. The effective coherence time is defined as $\tau = T_{1e}T_{2n}^*/(T_{1e}+T_{2n}^*)$, which accounts for both the electron spin relaxation ($T_{1e}$) and the nuclear spin dephasing ($T_{2n}^*$). The different colors represent varying overhead times $t_i$ required for initialization and readout processes. (b) The achievable sensitivity limits calculated at the optimal measurement times shown in (a). The results illustrate the dependence of the sensitivity on the system's coherence properties and the duty cycle constraints imposed by $t_i$.}
    \label{fig:5}
\end{figure}

Furthermore, a complete experimental cycle is not devoted exclusively to geometric-phase accumulation. 
State initialization (typically implemented through optical pumping and microwave transfer) and the final optical readout both require a non-negligible amount of time. 
We denote the total overhead associated with these operations by $t_i$. 
This overhead reduces the effective measurement duty cycle and introduces an additional sensitivity attenuation factor of $\sqrt{(t_i+T_m)/T_m}$. 
Taking both decoherence and timing overhead into account, the realistically achievable sensitivity $\eta_{\text{real}}$ is related to the ideal shot-noise limit $\eta_{\text{ideal}}$ by:
\begin{equation}
    \eta_{\text{real}} = \eta_{\text{ideal}} \exp\left(\frac{T_m}{\tau}\right)\sqrt{\frac{t_i+T_m}{T_m}}.
    \label{eq:realistic_sensitivity}
\end{equation}
Here we assume that decoherence reduces the signal contrast multiplicatively, so that the phase uncertainty is increased by the inverse attenuation factor.

To optimize the sensor performance, the measurement time $T_m$ must be chosen to minimize Eq.~(\ref{eq:realistic_sensitivity}) for a given set of system parameters, namely $\tau$ and $t_i$. 
Figure~\ref{fig:5}(a) shows the numerically obtained optimal measurement time as a function of the effective coherence time $\tau$. 
In the idealized case of instantaneous initialization and readout ($t_i = 0$ ms, blue dots), the optimal $T_m$ scales linearly with $\tau$. 
By contrast, as the overhead time $t_i$ increases (orange and green dots), the optimal measurement time shifts to larger values in order to mitigate the duty-cycle penalty.

The impact of these temporal constraints on the ultimate device performance is quantified in Fig.~\ref{fig:5}(b), which shows the achievable sensitivity evaluated at the corresponding optimal measurement time $T_m$. 
As expected, increasing the effective coherence time $\tau$ leads to a monotonic improvement in sensitivity. 
For a highly coherent diamond sample with $\tau\sim 10$ ms, the sensitivity approaches the $\mu	\text{rad/s}/\sqrt{\text{Hz}}$ regime, regime and becomes relatively insensitive to variations in the overhead time. 
By contrast, in more typical microdiamond platforms, where surface-charge fluctuations may limit $\tau$ to around $1$ ms, the overhead time $t_i$ becomes a critical factor. 
As illustrated in Fig.~\ref{fig:5}(b), increasing $t_i$ from $0$ ms to $1$ ms at $\tau \sim 1$ ms degrades the achievable sensitivity from approximately $3~\mu\text{rad/s}/\sqrt{\text{Hz}}$ to nearly $10~\mu\text{rad/s}/\sqrt{\text{Hz}}$.

Beyond spin decoherence, realistic implementations (particularly in levitated or high-speed rotating platforms) are also subject to imperfect fluorescence collection and mechanical frequency noise \cite{zhang2025ultra}. 
Finite photon-collection efficiency reduces the readout contrast and adds photon shot noise, while rotation-frequency jitter can produce additional random phase fluctuations.
Even so, provided that efficient optical interfaces and feedback stabilization are available, the dominant performance constraints remain the spin-coherence time and the protocol overhead identified in Fig.~\ref{fig:5}. 
This analysis therefore points to two primary experimental priorities: improving diamond material quality to extend $T_{1e}$ and $T_{2n}^*$, and reducing initialization/readout overhead through faster control protocols.

\section{conclusion}

We have proposed a high-precision rotation-sensing protocol based on the geometric phase of an NV-center ensemble. Operating in the near-resonant regime strongly enhances the phase response to external rotation, but also renders the dynamics vulnerable to nonadiabatic transitions near the avoided crossing. To overcome this limitation, we introduced a counter-diabatic control protocol derived from the Kato gauge potential, which suppresses nonadiabatic leakage while preserving the desired geometric-phase accumulation.

In the ideal shot-noise-limited case, our analysis predicts that an ensemble of $10^6$ active spins can achieve a rotation sensitivity of approximately $0.6~\mu\mathrm{rad/s}/\sqrt{\mathrm{Hz}}$. This corresponds to a four-orders-of-magnitude improvement over the conventional detuned protocol. When a finite effective coherence time $\tau$, determined jointly by $T_{1e}$ and $T_{2n}^*$, as well as the initialization and readout overhead $t_i$, are taken into account, sensitivities in the $\mu\mathrm{rad/s}/\sqrt{\mathrm{Hz}}$ range remain achievable after optimization of the interrogation time.

These results establish a realistic route toward geometric-phase-based solid-state quantum gyroscopes. By combining near-resonant enhancement with counter-diabatic protection, the proposed protocol provides a promising strategy for realizing ultra-compact quantum gyroscopes for inertial sensing and navigation.

\begin{acknowledgments}
This work is supported by the National Natural Science Foundation of
China (Grant No. 12441502) and the Beijing Institute of Technology Research Fund Program under Grant No. 2024CX01015. 
\end{acknowledgments}

\bibliography{main}
\end{document}